\documentclass[twocolumn,aps,prl,amssymb,amstext,showpacs,nobalancelastpage]{revtex4}
\usepackage{graphicx,float,amssymb,amstext,amsmath}% Include figure files
\usepackage{dcolumn}% Align table columns on decimal point
\usepackage{bm}% bold math

\newcommand{\rem}[1]{}

\newcommand{\new}[1]{{ #1}}

\newcommand{\R}{{\mathbb R}}
\newcommand{\Z}{{\mathbb Z}}

\def\dd{\mbox{d}}

\newcommand{\capsty}{\footnotesize}

\newcommand{\Capts}[1]{}
\newcommand{\FIGo}[3]{\begin{figure}%
#3% % comment this line out to remove all figures
\caption[]{\capsty #2}%
\label{#1}%
\end{figure}}

\begin{document}

\title{Non-uniqueness of phase shift in central scattering due to monodromy}

\author{Holger R. Dullin} %\email{hdullin@usyd.edu.au}
\affiliation{
School of Mathematics and Statistics, University of Sydney,
NSW 2006, Australia}

\author{Holger Waalkens} %\email{h.waalkens@rug.nl}
\affiliation{Department of Mathematics, University of Groningen, 9747 AG Groningen, the Netherlands,\\
and School of Mathematics, University of Bristol, Bristol BS8 1TW, UK}

\date{\today}

\begin{abstract}

Scattering at a central potential is completely characterized by the phase shifts which are the differences in phase between outgoing 
scattered and unscattered partial waves. In this letter it is shown that, for 2D scattering at a repulsive central potential, 
the phase shift cannot be uniquely  defined due to a topological obstruction which is similar to  {\em monodromy} in bound systems.

\end{abstract}

\pacs{45.05.+x, 03.65.Nk, 03.65.Sq, 45.40.-j}

\maketitle

{\em Introduction.---}
In physics it is often crucial to find ``suitable'' coordinates.  
One important set of coordinates in Hamiltonian mechanics are the action-angle variables \cite{Gold80}. 
Action-angle variables played a crucial role in the development of early quantum mechanics 
in the Bohr-Sommerfeld quantization rule. 
The existence of action-angle variables was already addressed by Einstein in 1917 \cite{Einstein17}. He argued that a quantization of actions only works if the mechanical system is integrable, i.e.\ it has as many independent constants of motion in involution as degrees of freedom. The existence of action-variables was made more precise in the  Liouville-Arnold theorem \cite{Arnold78}: if a connected component of the common level set of the constants of motion is regular (i.e.\ the gradients of the constants of motions are everywhere linearly independent on the component) and bounded then it has the topology of a torus and action-angle variables exist in the neighborhood of the torus. The angles are the coordinates on the individual tori (which are just Cartesian products of circles) and the actions change from torus to torus in a smooth way. 
This theorem is local near a regular torus. The question of if and how these ``local action-angle variables'' fit together \emph{globally} has been ignored for a long time. Recently it has been shown that there can exist topological obstructions to the global uniqueness of action-angle variables \cite{Duistermaat80}.
Consider a family of regular invariant tori in phase space that starts and ends with the same torus. 
This family may be a non-trivial torus bundle. Similarly, a M\"obius strip is a
non-trivial interval bundle: even though every local piece of it is just a rectangle, globally 
it is twisted. If the action variables are changed by a non-trivial uni-modular transformation 
after one (mono) circuit (dromos) through a family of regular invariant tori, then the system has \emph{monodromy}.
Monodromy implies that the action-angle variables do {\em not} give global coordinates.
It is a topological obstruction, because the twist in the bundle cannot be removed by smooth 
deformations of the bundle. 
By contrast, if the loop of regular tori can be contracted (passing only through regular tori)
it cannot have monodromy. In this way monodromy is related to a non-regular level set 
of the constants of motion that is not a torus. Such critical sets appear in phase space where 
the gradients of the constants of motion are linearly dependent.
The most prominent example of a critical set that causes monodromy
is a pinched torus \cite{CushBates}. It exists in integrable systems with two degrees of freedom that have  an 
unstable equilibrium point of focus-focus type,
i.e.\ with eigenvalues of the form $\alpha \pm i \omega$, $-\alpha \pm i \omega$, $\alpha \not = 0$. 
One of the simplest examples with this type of monodromy is the spherical pendulum \cite{CushBates}.
The quantum version of this phenomenon \cite{CushDuist88} explains why there is no global 
quantum number assignment
for the hydrogen atom in external fields \cite{CushSad99},
the H$_2^+$ molecular ion \cite{WaalkensDullinRichter04},
the rovibrational spectrum of CO$_2$ \cite{CDGHJLSZ04}, and other systems \cite{SadovskiiZhilinskii99}.

In this letter we study the implication of the unbounded analog of monodromy in scattering problems.
Elastic scattering at a central potential is completely characterized by the phase shifts which are the differences in phase of outgoing 
{\em scattered} partial waves 
and outgoing
{\em unscattered} partial waves.
For a planar system,
a partial wave $ \langle x,y | l,p\rangle $ with angular momentum $l$ and asymptotic momentum $p$
at infinity 
gains a phase $\delta(l,p)$. 
In fact, the action of the scattering matrix $S$ on a partial wave  is
\begin{equation} \label{eqn:Smat}
S|l,p\rangle = \exp\big( 2i\delta(l,p) \big) |l,p\rangle \,.
\end{equation}
All physical quantities such as  scattering cross sections and amplitudes can be expressed in terms of $\delta(l,p)$.

The phase shift is positive for attractive potentials and negative for repulsive potentials.
If there is no interaction then $\delta(l,p)=0$.
Similarly, the phase shift vanishes in the limiting case of large $p$ where the potential can be ignored due to the dominating kinetic energy.
The common procedure to define the phase shift therefore is to smoothly continue $\delta(l,p)$ from large to small $p$. 
We will show that for smooth repulsive potentials, there is a topological obstruction to this procedure, 
and as a consequence, the phase shift cannot be uniquely defined.

\goodbreak

{\em Nonuniqueness of phase shift.---}
\new{
Consider a smooth repulsive central potential $V(r)$
with $V(r) \to 0$ sufficiently fast for $r\to \infty$.  The Taylor expansion
at the origin is $V(r) = E_c - \mu \alpha^2 r^2/2 + {\cal O}(r^4) $ with $E_c > 0$.
}
A semiclassical expression for $\delta(l,p)$ can be obtained from the WKB method \cite{FF65}. 
Assuming $\delta(l,p)\rightarrow 0$ for $p \rightarrow \infty$ the WKB approximation yields
$\delta_{\text{WKB}}(l,p)=\Delta W(l,p)/(2\hbar)$ where $\Delta W(l,p)$ is the difference of the radial actions 
with and without potential,
\begin{align}
\label{eq:defW}
& \Delta W (l,p) =  W(l,p)-W'(l,p)  :=  \\ 
& 2\int_{r_0}^\infty \sqrt{p^2 - l^2/r^2 - 2\mu V(r)} \, \dd r 
 - 2\int_{r'_0}^\infty \sqrt{p^2 - l^2/r^2 } \, \dd r\,. \nonumber
\end{align}
Here $r_0$ and $r'_0$ are the classical turning points with and without potential, i.e., 
$r_0$ is the largest nonnegative root of $r^2 p^2 - l^2 - 2\mu r^2 V(r)$ or zero if $l=0$ in combination with $p>p_c=(2\mu E_c)^{1/2}$, and $r'_0=|l|/p$.  
The difference $\Delta W$ is finite while the 
individual integrals diverge (see Fig.~\ref{fig:phaseportrait}).

\def\figphaseportrait{%
Phase portrait $(r,p_r)$ with $p_r^2 = 2\mu E - l^2/r^2 - 2\mu V(r)$ with $E=3$, $l=1$, and $V(r)\equiv 0$ (outer curve) and $V(r) = a/(1+(br)^2)$ with $a=20$ and $b=1$
(inner curve). The shaded area is equal to $\Delta W$ defined in (\ref{eq:defW}). 
}
\def\FIGphaseportrait{
\begin{center} % /home/mazhw/Cpp/ScatteringMonodromy/phase_portrait
\includegraphics[width=6.5cm]{fig_1}
\end{center}
}
\FIGo{fig:phaseportrait}{\figphaseportrait}{\FIGphaseportrait}

Surprisingly the function $\Delta W(l,p)$ is not globally smooth: 
it is not differentiable at $l=0$ when $p < p_c$.
To illustrate this we show contours of $\Delta W$ 
in Fig.~\ref{fig:phaseshift}a.
Consider the derivative of $\Delta W$ with respect to $l$,
\begin{align}
\label{eq:dWdL}
&\frac{\partial }{\partial l} \Delta W (l,p) = \frac{\partial W(l,p)}{\partial l} - \frac{\partial W'(l,p)}{\partial l} =\\
& l \int_{z_0}^\infty \frac{-1}{z\sqrt{ z p^2 - l^2 - 2\mu z U(z)} } \, \dd z 
  - l \int_{z'_0}^\infty \frac{-1}{z\sqrt{z p^2 - l^2} } \, \dd z\,. \nonumber
\end{align}
Here we substituted $z=r^2$ and let $U(z) \equiv V(\sqrt{z})$. 
In contrast to the integrals in (\ref{eq:defW}) their derivatives with respect to $l$ exist individually. 
The second integral is elementary and gives
$ -\text{sgn}(l) \pi$. 

For general $l$ and $p$, the integral $\partial W(l,p)/\partial l$ in (\ref{eq:dWdL}) depends on the potential $U$. 
Interestingly, for $l=0$, this is no longer the case.
The limiting case $l\rightarrow 0$ is tricky: 
depending on whether $p>p_c$ or $p<p_c$ the branch point  $z_0$ of the square root in the 
integrand either collides or does not collide with the integrand's pole at $z=0$ as 
$l\rightarrow 0$. 
The collision of the branch point and the pole leads to the divergence of the integral and the question arises
of how this divergence is compensated by the vanishing of the prefactor $l$.
At $z=0$ the argument of the square root has the Taylor expansion
\begin{equation} \label{eq:Taylorsqrt}
-l^2 + (p^2-p_c^2)z + \mu^2 \alpha^2 z^2 + {\cal O}(z^3)\,. 
\end{equation}
\new{ We note that in case $\alpha=0$
higher order terms can be included and do not lead to a substantial change of the following argument.
}
For $p<p_c$, the linear term of this expression has a negative coefficient that does not depend on $l$. 
Hence, the collision of $z_0$ and zero as $l\rightarrow 0$
does not take place,
and the integral is not critical in this limiting case. 
Due to the prefactor $l$ in (\ref{eq:dWdL}) we thus have $\partial W(l,p)/\partial l\rightarrow 0$ as $l\rightarrow 0$ and accordingly, the left and right hand derivatives of $\Delta W(l,p)$ with respect to $l$ at zero are
\begin{equation}
\lim_{l\rightarrow 0\mp}\frac{\partial \Delta W(l,p)}{\partial l} = \mp \pi  \quad (p<p_c)\,.
\end{equation}
\def\figphaseshift{%
$(l,p)$-plane with contours of (a) $\Delta W$ defined in (\ref{eq:defW}) and (b) $\Delta \tilde{W}$ defined in (\ref{eqn:smoothed}).
The bold dot marks $(l,p)=(0,p_c)$. 
The potential is the same as in Fig.~\ref{fig:phaseportrait}.
}
\def\FIGphaseshift{
\centerline{
\raisebox{4.5cm}{a)}%/home/mazhw/Cpp/ScatteringMonodromy/phase_shift_conts
\includegraphics[width=7.0cm]{fig_2a}
}
\centerline{
\raisebox{4.5cm}{b)}%/home/mazhw/Cpp/ScatteringMonodromy/phase_shift_prime_conts
\includegraphics[width=7.0cm]{fig_2b}
}
}
\FIGo{fig:phaseshift}{\figphaseshift}{\FIGphaseshift}

\def\figfig2{%
Complex $z$-plane with integration paths for the differential $-\dd z/(2 z (z p^2 - l^2 - 2\mu z U(z))^{1/2} ) $ which has a pole at the origin 
$z=0$ (marked by the cross). 
The square root is real along the branch cut which extends from the turning point $z_0$ (marked by the dot) along the postive real axis; 
it is positive `above' and negative `below' this branch cut. The integration path $C$ (solid line) is equivalent to the composition
$C = C_1+C_2$ (dashed line).
}
\def\FIGfig2{
\begin{center}%Figures/r_plane
\includegraphics[width=7.0cm]{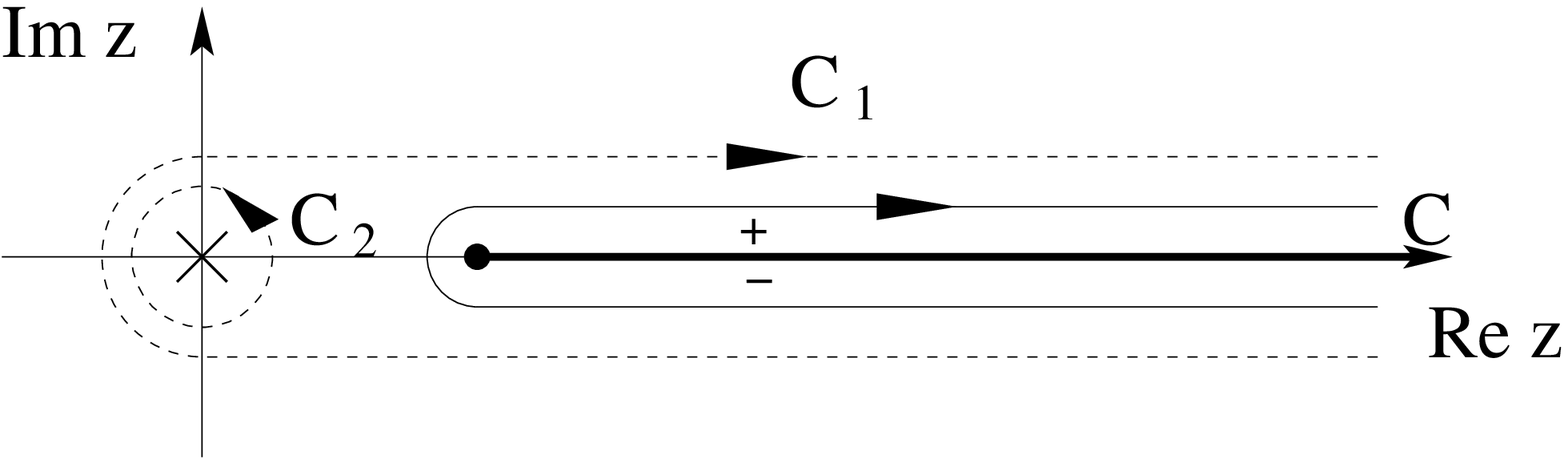}
\end{center}
}
\FIGo{fig:fig2}{\figfig2}{\FIGfig2}

When $p>p_c$ the coefficient of the linear term in (\ref{eq:Taylorsqrt}) is positive and the collision of the branch point and the pole {\em does}  take place.
Consider the integral  in the complex plane. For $l\neq 0$ we define the integration path $C$ as shown in Fig.~\ref{fig:fig2}. 
This gives
\begin{equation} \label{eq:wbyl_C}
\frac{\partial W(l,p)}{\partial l} =
l \int_C \frac{-1}{2 z \sqrt{ z p^2 - l^2 - 2\mu z U(z)} } \, \dd z \,.
\end{equation}
To study the limit $l\rightarrow 0$ we deform the integration path $C$ by 
wrapping it over the pole at zero and compensate the capture of the pole by adding a small closed 
integration path that encircles the pole in opposite 
direction.
Thus we decompose $C=C_1 + C_2$ with $C_1$ and $C_2$ as shown in Fig.~\ref{fig:fig2}. 
The integal (\ref{eq:wbyl_C}) thus becomes 
\begin{equation} \label{eq:wbyl_C1C2}
\frac{\partial W(l,p)}{\partial l} = \sum_{k=1}^2
l \int_{C_k} \frac{-1}{2 z \sqrt{ z p^2 - l^2 - 2\mu z U(z)} } \, \dd z \,.
\end{equation}
The integration path $C_1$ is not critical for $l\rightarrow 0$ and due to the prefactor $l$
the contribution to $\partial W/ \partial l$ vanishes for $l\rightarrow 0$.
For the choice of the branch of the square root explained in Fig.~\ref{fig:fig2} 
the differential $-\dd z/(2z\sqrt{ z p^2 - l^2 - 2\mu z U(z)})$ 
has residue  $-i/(2|l|)$ at $z=0$. The integral along $C_2$ thus leads to the contribution
\begin{equation} 
l \int_{C_2} \frac{-1}{2 z \sqrt{ z p^2 - l^2 - 2\mu z U(z)} } \, \dd z = \text{sgn}(l) \pi \,.
\end{equation}

We thus find that for \new{$p>p_c$}, the contribution of $\partial W(l,p)/\partial l$ and $\partial  W'(l,p)/\partial l$ to  $\partial \Delta W(l,p)/\partial l$ cancel each other for $l\rightarrow 0$ and accordingly,
\begin{equation}
\lim_{
l\rightarrow0
}\frac{\partial }{\partial l} \Delta W (l,p) = 0 \quad (p>p_c) \,.
\end{equation}

One might think of removing the kink in Fig.~\ref{fig:phaseshift}a by `smoothing' $\Delta W (l,p)$ according to
\begin{equation} \label{eqn:smoothed}
\Delta \tilde{W} (l,p) =
\left\{ 
\begin{array}{cl}
\Delta W (l,p)          & \mbox{for }l \le 0 \\ 
\Delta W (l,p)  - 2\pi l & \mbox{for }l>0  
\end{array}
\right. \,.
\end{equation}
This however introduces a kink at the segment of $l=0$ where $p>p_c$, see Fig.~\ref{fig:phaseshift}b.

The WKB method gives $2 \hbar \delta = \Delta W$ where $l = m \hbar$, $m \in \Z$.
The values of $\delta$ are only relevant $\bmod \, \pi$, see \eqref{eqn:Smat}.
In other words, the function $\exp(2 i \delta)$ is (locally) periodic.
In order to study this periodicity we consider the values $p = k \hbar$ such that
$\delta( m \hbar, k \hbar) = 0 \bmod \pi$, see Fig.~\ref{fig:fig1}.
The function $\exp(2 i \delta)$ is not globally periodic because of the singularity 
at $(m,k) = (0, p_c/\hbar)$. This can be seen by transporting a unit cell 
in the lattice around the singularity. 
The lattice cell crosses the line $m=0$ according to the modified $\Delta \tilde W$ 
of \eqref{eqn:smoothed} which is smooth for $p < p_c$, 
while the original $\Delta W$ is smooth for $p > p_c$.
Thus in the presence of a repulsive localised potential the phase shift $\delta$ 
cannot be globally defined. 
We call this phenomenon \emph{quantum scattering monodromy}.

\def\figfig1{%
Lattice of zeros (empty circles)  of the 
phase shift $\delta$ mod $\pi$ in the plane $m=l/\hbar$, $k=p/\hbar$, and parallel transport of a lattice cell about the singularity $(m,k)=(0,p_c/\hbar)$ (filled circle). 
The potential is the same as in Fig.~\ref{fig:phaseportrait}. $\hbar=0.25$.
}
\def\FIGfig1{
\begin{center}%/home/mazhw/Cpp/ScatteringMonodromy/phase_shift_spectr_xfig
\includegraphics[width=7.0cm]{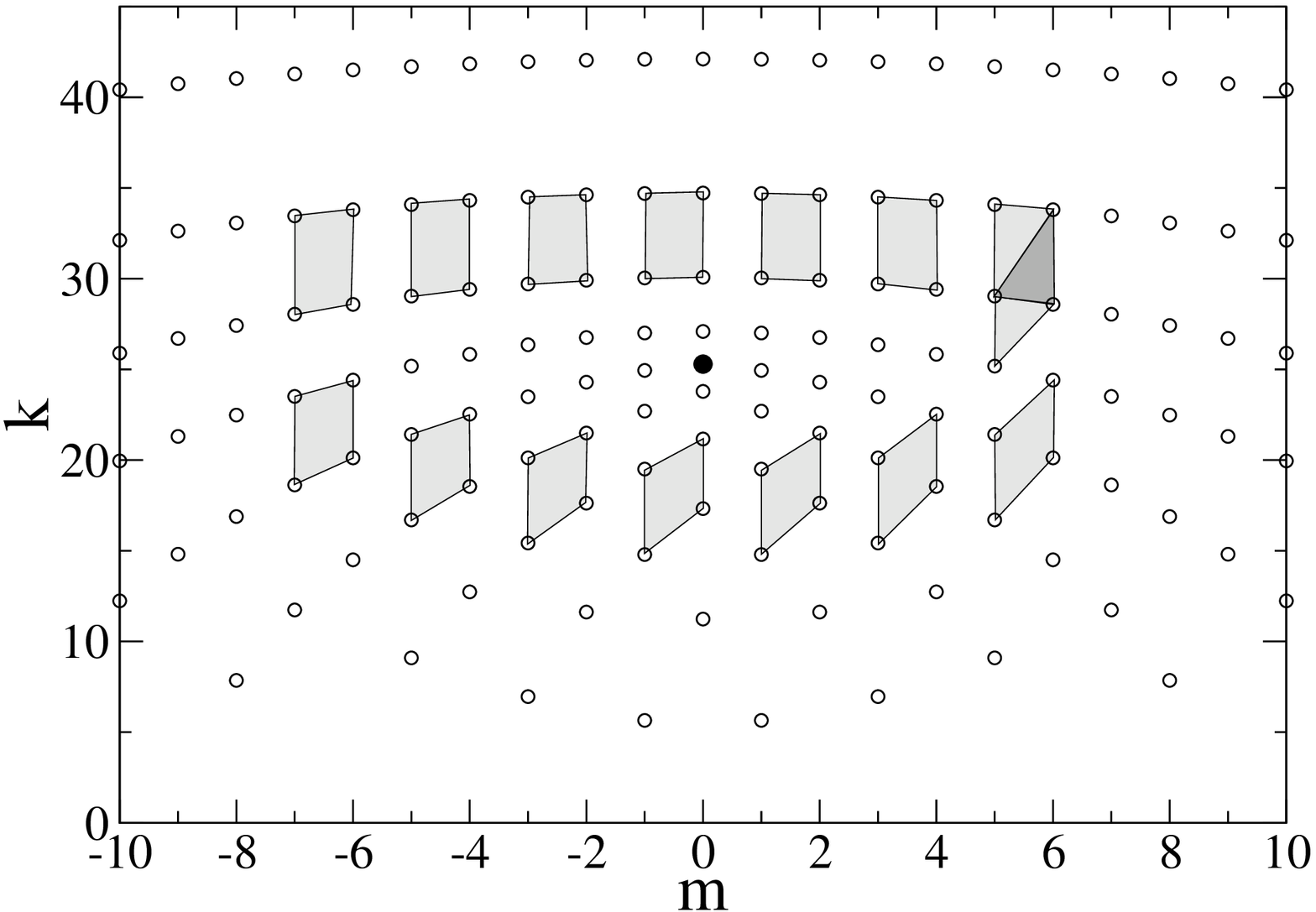}
\end{center}
}
\FIGo{fig:fig1}{\figfig1}{\FIGfig1}

Notice that the derivative of the phase shift with respect to the energy 
give the eigenvalues of the   
Wigner-Smith time delay matrix $Q=-i\hbar S^{-1} \partial S/\partial E$. 
Semiclassically, this derivative is given by
$\partial \Delta W(l,p)/\partial E=\Delta T$ which is the classical time delay.
This derivative is smooth everywhere apart from the point $(l,E)=(0,E_c)$.

The WKB approximation (\ref{eq:defW}) does not  account for the collision of the classical turning 
point and the singularity of the effective potential when $l \to 0$.
An asymptotic expansion of the exact solution of the
radial wave equation shows that nevertheless the error is less than 1\%.

{\em Classical explanation.---}
The classical interpretation of $\partial \Delta W (l,p) / \partial l$ is the angle of deflection.
Consider the polar angle $\varphi$ between the incoming and outgoing orbit in configuration space,
\begin{equation}
 \varphi = \int_{-\infty}^{\infty} \dot{\varphi} \,\dd t = 
 \int_{-\infty}^{\infty} \frac{l}{\mu r^2} \,\dd t\,.
\end{equation}
Substituting $\dd t = \dd r/\dot{r}$ gives
\begin{eqnarray}
\label{eq:deltavarphi}
 \varphi &=& \int_\infty^{r_0} \frac{l}{\mu r^2} \frac{1}{\dot{r}} \,\dd r +
\int_{r'_0}^\infty \frac{l}{\mu r^2} \frac{1}{\dot{r}} \,\dd r   \\
&=&2 \int_{r_0}^\infty  \frac{l}{r\sqrt{2\mu Er^2-l^2-2\mu r^2 V(r)}} \,\dd r
\end{eqnarray}
where $r_0$ is the turning point.
We used $\dot{r}<0$ in the first integral and $\dot{r}>0$ in the second integral in (\ref{eq:deltavarphi}).
Up to the sign
the derivative $\partial \Delta W (l,p)/\partial l$ in (\ref{eq:dWdL}) thus coincides with the angle of deflection 
$\Delta \varphi$ between the scattered and the corresponding unscattered orbit.

\def\figmonconf{%
Orbits in configuration space coming in from $y=-\infty$ for four different pairs of angular momenta  
and asymptotic momenta 
marked as bold  points on the path in the $(l,p)$-plane that encircles the critical point $(l,p)=(0,p_c)$. 
}
\def\FIGmonconf{
\begin{center}%Figures/monodr
\includegraphics[width=7.0cm]{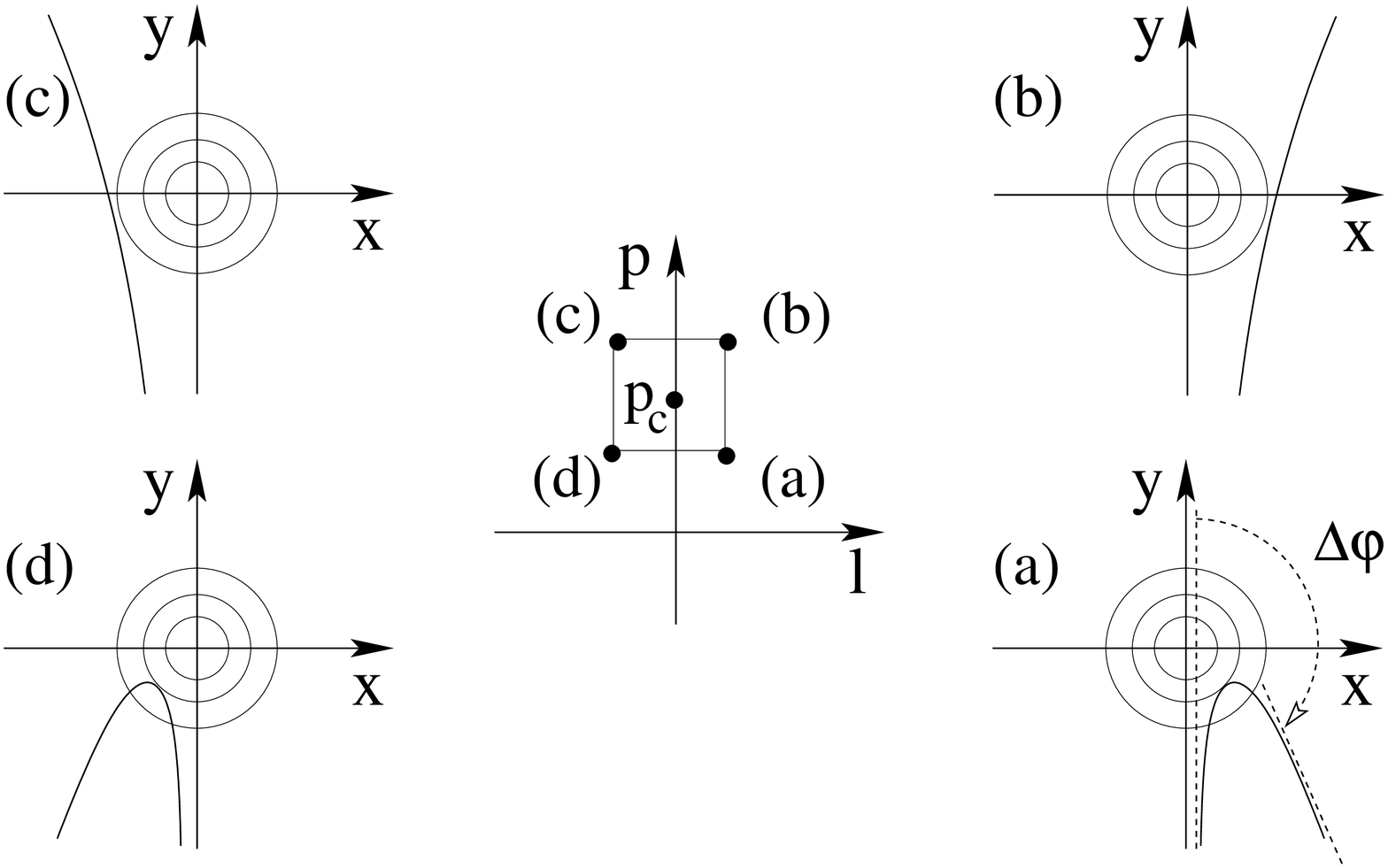}
\end{center}
}
\FIGo{fig:monconf}{\figmonconf}{\FIGmonconf}

Let us now follow the deflection angle $\Delta \varphi$ for pairs of angular momenta 
(or equivalently impact parameters) and asymptotic momenta (i.e.\ energies) 
along a closed path in the $(l,p)$-plane that encircles the critical point $(l,p)=(0,p_c)$, 
see~Fig.~\ref{fig:monconf}.
Consider orbits that come in from $y=-\infty$. %  as shown in Fig.~\ref{fig:monconf}.
For $l=0$ and $p<p_c$, the particle comes in along the $y$-axis and
slows down. Due to insufficient energy it cannot overcome 
the potential barrier and so turns back towards $y=-\infty$. This orbit has $\Delta \varphi =\pi$. 
If $l$ is increased  to a value $l=l_{\text{(a)}}>0$ (keeping $p<p_c$ fixed) 
then the particle gets deflected to the right 
and $\Delta \varphi$ decreases to a value $\pi>\Delta \varphi_{\text{(a)}}>0$. 
If we now increase $p$ to a value $p_{\text{(b)}}>p_c$ 
(keeping $l$ fixed) the deflection angle $\Delta \varphi$ decreases to a value $0<\Delta \varphi_{\text{(b)}}<\Delta \varphi_{\text{(a)}}$.
If we then decrease $l$  (keeping $p$ fixed) 
the deflection angle decreases further. At $l=0$ the particle comes in along the $y$-axis, slows down, but now has sufficient energy to cross the barrier and move towards $y=+\infty$. This orbit has $\Delta \varphi =0$.
If $l$ is decreased further to a negative value $l_{\text{(c)}}<0$ the particle gets deflected to the left giving a negative deflection angle $-\pi<\Delta \varphi_{\text{(c)}}<0$. If $p$ is then decreased to a value $p_{\text{(d)}}<p_c$ (keeping $l$ at $l_{\text{(c)}}<0$) the defection angle decreases further to a value $-\pi<\Delta \varphi_{\text{(d)}}<\Delta \varphi_{\text{(c)}}$. If $l$ is increased (keeping $p$ fixed at $p_{\text{(d)}}$) the deflection angle decreases even further and at $l=0$ it reaches the value $\Delta \varphi = -\pi$. Upon returning to the starting point of closed path $\gamma$ in the $(l,p)$-plane that encircles the critical point $(l,p)=(0,p_c)$ the deflection angle thus is increased by $2\pi$,
\begin{equation}
\oint_\gamma \frac{\partial \Delta W(l,p)}{\partial l}\mbox{d}l + \frac{\partial \Delta W(l,p)}{\partial p}\mbox{d}p = 2\pi\,.
\end{equation}
Such a non-zero value from a closed loop $\gamma$ only occurs when the critical point 
$(0, p_c)$ is encircled by $\gamma$.
If a system has loops of regular values for which the deflection angle is increased by (multiples of) $2\pi$ 
we say the system has  \emph{scattering monodromy}.

For the classical system, the angular momentum $L$ and
the Hamiltonian function $H$ are two independent constants of motion which are in involution.
The classical system is therefore integrable.
The level set  $\{ L = l, H = E \}$ in phase space
for a regular value $(l,E)$ topologically is a cylinder.  
The invariant cylinder $\R \times S^1$ consists of an orbit ($\sim \R$) as shown in Fig.~\ref{fig:monconf} 
and all its partners with different angle of incidence ($\sim S^1$) but the same $l$ and $E$.
At the critical value $(0, E_c)$ the gradients of $L$ and $H$ are linearly dependent. 
The critcial level set $\{ L = 0, H = E_c \}$ is topologically a cone. 
It consists of the equilibrium point at the origin and all orbits approaching it forward
or backward in time. 
The invariant cone  (a pinched cylinder) in the phase space of a scattering system is the analogue of the 
pinched torus in a bound system. 
A loop of invariant cylinders that encircles the invariant cone cannot be contracted, 
and as a result $\Delta \varphi$ shows scattering monodromy.

{\em Conclusions.---}
We have shown that the quantum scattering phase shift $\delta$ for a smooth radially symmetric 
repulsive potential % that decays sufficiently fast at infinity 
cannot be globally defined. 
The classical analogue is that the deflection angle changes by $2\pi$ upon traversing a 
loop in the space of constants of motion
that encloses the critical value corresponding to the equilibrium point. 
\new{As opposed to the more abstract consequences of monodromy in compact systems
this phenomenon is `directly' observable, e.g. by playing marbles (neglecting moments of inertia) on a surface with a rotationally symmetric bump. 
}


\begin{thebibliography}{11}
\expandafter\ifx\csname natexlab\endcsname\relax\def\natexlab#1{#1}\fi
\expandafter\ifx\csname bibnamefont\endcsname\relax
  \def\bibnamefont#1{#1}\fi
\expandafter\ifx\csname bibfnamefont\endcsname\relax
  \def\bibfnamefont#1{#1}\fi
\expandafter\ifx\csname citenamefont\endcsname\relax
  \def\citenamefont#1{#1}\fi
\expandafter\ifx\csname url\endcsname\relax
  \def\url#1{\texttt{#1}}\fi
\expandafter\ifx\csname urlprefix\endcsname\relax\def\urlprefix{URL }\fi
\providecommand{\bibinfo}[2]{#2}
\providecommand{\eprint}[2][]{\url{#2}}

\bibitem[{\citenamefont{Goldstein}(1980)}]{Gold80}
\bibinfo{author}{\bibfnamefont{H.}~\bibnamefont{Goldstein}},
  \emph{\bibinfo{title}{Classical Mechanics}}
  (\bibinfo{publisher}{Addison-Wesley}, \bibinfo{address}{Reading, MA},
  \bibinfo{year}{1980}), \bibinfo{edition}{2nd} ed.

\bibitem[{\citenamefont{Einstein}(1917)}]{Einstein17}
\bibinfo{author}{\bibfnamefont{A.}~\bibnamefont{Einstein}},
  \bibinfo{journal}{Verh. DPG} \textbf{\bibinfo{volume}{19}},
  \bibinfo{pages}{82} (\bibinfo{year}{1917}).

\bibitem[{\citenamefont{Arnold}(1978)}]{Arnold78}
\bibinfo{author}{\bibfnamefont{V.~I.} \bibnamefont{Arnold}},
  \emph{\bibinfo{title}{Mathematical Methods of Classical Mechanics}},
  vol.~\bibinfo{volume}{60} of \emph{\bibinfo{series}{Graduate Texts in
  Mathematics}} (\bibinfo{publisher}{Springer}, \bibinfo{address}{Berlin},
  \bibinfo{year}{1978}).

\bibitem[{\citenamefont{Duistermaat}(1980)}]{Duistermaat80}
\bibinfo{author}{\bibfnamefont{J.~J.} \bibnamefont{Duistermaat}},
  \bibinfo{journal}{Comm. Pure Appl. Math.} \textbf{\bibinfo{volume}{33}},
  \bibinfo{pages}{687} (\bibinfo{year}{1980}).

\bibitem[{\citenamefont{Cushman and Bates}(1997)}]{CushBates}
\bibinfo{author}{\bibfnamefont{R.~H.} \bibnamefont{Cushman}} \bibnamefont{and}
  \bibinfo{author}{\bibfnamefont{L.~M.} \bibnamefont{Bates}},
  \emph{\bibinfo{title}{Global Aspects of Classical Integrable Systems}}
  (\bibinfo{publisher}{Birkh{\"a}user}, \bibinfo{address}{Basel, Boston,
  Berlin}, \bibinfo{year}{1997}).

\bibitem[{\citenamefont{Cushman and Duistermaat}(1988)}]{CushDuist88}
\bibinfo{author}{\bibfnamefont{R.~H.} \bibnamefont{Cushman}} \bibnamefont{and}
  \bibinfo{author}{\bibfnamefont{J.~J.} \bibnamefont{Duistermaat}},
  \bibinfo{journal}{Bull. Amer. Math. Soc.} \textbf{\bibinfo{volume}{19}},
  \bibinfo{pages}{475} (\bibinfo{year}{1988}).

\bibitem[{\citenamefont{Cushman and Sadovskii}(1999)}]{CushSad99}
\bibinfo{author}{\bibfnamefont{R.~H.} \bibnamefont{Cushman}} \bibnamefont{and}
  \bibinfo{author}{\bibfnamefont{D.~A.} \bibnamefont{Sadovskii}},
  \bibinfo{journal}{Europhys. Lett.} \textbf{\bibinfo{volume}{47}},
  \bibinfo{pages}{1} (\bibinfo{year}{1999}).

\bibitem[{\citenamefont{Waalkens et~al.}(2004)\citenamefont{Waalkens, Dullin,
  and Richter}}]{WaalkensDullinRichter04}
\bibinfo{author}{\bibfnamefont{H.}~\bibnamefont{Waalkens}},
  \bibinfo{author}{\bibfnamefont{H.~R.} \bibnamefont{Dullin}},
  \bibnamefont{and} \bibinfo{author}{\bibfnamefont{P.~H.}
  \bibnamefont{Richter}}, \bibinfo{journal}{Physica D}
  \textbf{\bibinfo{volume}{196}}, \bibinfo{pages}{265} (\bibinfo{year}{2004}).

\bibitem[{\citenamefont{Cushman et~al.}(2004)\citenamefont{Cushman, Dullin,
  Giacobbe, Holm, Joyeux, Lynch, Sadovskii, and Zhilinskii}}]{CDGHJLSZ04}
\bibinfo{author}{\bibfnamefont{R.~H.} \bibnamefont{Cushman}},
  \bibinfo{author}{\bibfnamefont{H.~R.} \bibnamefont{Dullin}},
  \bibinfo{author}{\bibfnamefont{A.}~\bibnamefont{Giacobbe}},
  \bibinfo{author}{\bibfnamefont{D.~D.} \bibnamefont{Holm}},
  \bibinfo{author}{\bibfnamefont{M.}~\bibnamefont{Joyeux}},
  \bibinfo{author}{\bibfnamefont{P.}~\bibnamefont{Lynch}},
  \bibinfo{author}{\bibfnamefont{D.~A.} \bibnamefont{Sadovskii}},
  \bibnamefont{and}
  \bibinfo{author}{\bibfnamefont{B.}~\bibnamefont{Zhilinskii}},
  \bibinfo{journal}{Phys. Rev. Lett.} \textbf{\bibinfo{volume}{93}},
  \bibinfo{pages}{024302} (\bibinfo{year}{2004}).

\bibitem[{\citenamefont{Sadovski{\'{\i}} and
  Z{\^h}ilinski{\'{\i}}}(1999)}]{SadovskiiZhilinskii99}
\bibinfo{author}{\bibfnamefont{D.~A.} \bibnamefont{Sadovski{\'{\i}}}}
  \bibnamefont{and} \bibinfo{author}{\bibfnamefont{B.~I.}
  \bibnamefont{Z{\^h}ilinski{\'{\i}}}}, \bibinfo{journal}{Phys. Lett. A}
  \textbf{\bibinfo{volume}{256}}, \bibinfo{pages}{235} (\bibinfo{year}{1999}).
  

\bibitem[{\citenamefont{Fr{\"o}man and Fr{\"o}man}(1965)}]{FF65}
\bibinfo{author}{\bibfnamefont{N.}~\bibnamefont{Fr{\"o}man}} \bibnamefont{and}
  \bibinfo{author}{\bibfnamefont{P.}~\bibnamefont{Fr{\"o}man}},
  \emph{\bibinfo{title}{JWKB Approximation: Contributions to the Theory}}
  (\bibinfo{publisher}{North-Holland Publishing Company},
  \bibinfo{address}{Amsterdam}, \bibinfo{year}{1965}).

\end{thebibliography}
\end{document}